# Coherent Absorption Synergizes with Plasmon-Enhanced Graphene Terahertz Photo-thermoelectric Response

*Runli Li†, Shaojing Liu†, Ximiao Wang, Hongjia Zhu, Yongsheng Zhu, Shangdong Li, and Huanjun Chen\**

State Key Laboratory of Optoelectronic Materials and Technologies, Guangdong Province Key Laboratory of Display Material and Technology, and School of Electronics and Information Technology, Sun Yat-sen University, Guangzhou 510275, China.

Corresponding authors: chenhj8@mail.sysu.edu.cn.

†These authors contributed equally.

ABSTRACT. Terahertz (THz) technology shows great potential in 6G communications and imaging, but faces challenges related to detector sensitivity, noise, and cryogenic operation. Here, we integrate interferometric enhancement of absorption (IEA) from a metal reflection layer with a graphene plasmon polariton atomic cavity (PPAC)-based photodetector. The hybrid configuration enhances the in-plane electric field and improves the plasmon-induced thermal gradient. Numerical simulations and photoresponse measurements were employed to systematically investigate the influence of a metal reflective layer on the photothermoelectric behavior of the device, which reveals the IEA design significantly boosts the THz absorption rate in graphene nanostructures and promotes asymmetry in the lateral diffusion of hot carriers. Compared with the bare device, the responsivity of the device is enhanced by approximately 30-folds, while maintaining a response time below 130 μs. We further demonstrate the potential of the device to distinguish concealed liquids, advancing high-responsivity, room-temperature, and compact terahertz imaging technology.

Terahertz waves, occupying a unique region of the electromagnetic spectrum between microwaves and infrared light, exhibit great potential in applications such as non-destructive imaging, high speed wireless communications, biomedical sensing, and security screening. [1-6] However, the development of high-performance terahertz detectors remains challenging due to the markedly reduced efficiency of conventional electronic and photonic devices in this frequency range[7]. Existing solutions, including Golay cells, bolometers, pyroelectric detectors, hot-carrier-based detectors, and field-effect transistors, generally rely on large effective areas, antenna-assisted focusing of terahertz radiation, or cooling atmosphere to suppress noise, and therefore suffer from limitations such as slow response, low quantum efficiency, narrow bandwidth, or the need for cryogenic operation. [7, 8] In recent years, two dimensional materials, especially graphene, have provided new ideas for overcoming this challenge due to their advantages of ultra-high carrier mobility, broadband absorption characteristics, and tunable Fermi level. [9, 10] Researchers have exploited two-dimensional semimetals (such as graphene, $WTe_2$, $PtTe_2$, and $T_d$-$MoTe_2$) and topological insulators (including $Bi_2Se_3$ and $PtSe_2$), in conjunction with integrated terahertz antennas, to realize room-temperature terahertz detection. [11, 12] However, these detectors still suffer from inherent drawbacks, including bulky device architectures, limited sensitivity, and narrow spectral bandwidth. Overcoming these challenges is crucial for the development of highly sensitive, fast-response, and miniaturized room-temperature terahertz detectors.

Graphene detectors based on the photothermoelectric (PTE) effect have attracted considerable attention in the scientific community due to their distinct advantages, including zero-bias operation, low power consumption, and ultrafast response.[13] When an asymmetric optical field or structural distribution induces a temperature gradient in graphene under illumination, the Seebeck effect drives the directional movement of hot carriers, thereby generating a photocurrent.[14] This mechanism is particularly suitable for the terahertz band, where the relatively low photon energy makes it challenging to efficiently generate electron–hole pairs via conventional photovoltaic mechanisms. The asymmetric electrode architecture, combined with a shortened graphene channel, enables the device to maintain a sub-nanosecond response time.[15] Moreover, resonant gate engineering allows flexible tuning of both the detection wavelength and the efficiency.[16] Theoretical studies have demonstrated that the linear Dirac cone band structure of graphene confers excellent hot carrier generation and transport properties within the terahertz regime.[17] Nevertheless, the inherently weak light-matter interaction, which arises from the atomic thickness of graphene, imposes a fundamental limitation on its photoconversion efficiency.[18] How to achieve antenna-free, high-responsivity graphene-based room-temperature terahertz detectors through structural design and innovation in physical mechanisms has emerged as a critical and unresolved challenge in this field.[19]

To strengthen the interaction between terahertz waves and graphene, various light localization strategies have been proposed, including plasmonic resonances,[20-22] Fabry–Pérot cavities,[23-26] and metasurfaces.[26-28] Among these, graphene plasmons represent a key mechanism for electromagnetic mode modulation that enables strong wave localization and field enhancement at deep subwavelength scales through the excitation of collective oscillations of free carriers in graphene.[29] Compared with conventional metal plasmons, graphene plasmons offer highly tunable resonant characteristics: their resonance frequency can be adjusted by modifying the dimensions, geometry, and periodicity of microstructures, and more importantly, can be dynamically tuned via gate voltage control of the Fermi level. This tunability enables resonant absorption enhancement, strong field confinement, and flexible spectral matching in the terahertz band, significantly improving detection sensitivity and spectral selectivity.

Metal reflectors have gained popularity due to their process compatibility and effective light manipulation capabilities. Thin films of gold or aluminum can reflect incident light to form a standing wave within dielectric layers, thereby extending the optical path and enhancing absorption. Combining graphene plasmonic structures with metal reflectors offers a promising route to further strengthen the interaction between terahertz waves and graphene: the coherent standing wave provided by the metal reflector can couple with graphene plasmon modes, yielding multiplicative field enhancement and thus markedly improving device responsivity. However, research in the terahertz band remains in its early stages. In particular, how the spacing between the metal reflector and graphene affects coherent superposition remains unclear, and the coupling between coherent optical fields and device responsivity has not been systematically explored.

In this study, we propose a novel detector architecture based on a coherent absorption enhanced graphene terahertz photodetector. The core innovation lies in the synergistic integration of interferometric enhancement of absorption (IEA) with a plasmon polariton atomic cavity (PPAC) on graphene.[30] The IEA structure utilizes a back-side metal reflector to create a standing wave through coherent superposition of incident and reflected terahertz waves. By precisely engineering the spacing between the metal reflector and the graphene microstructure, we achieve constructive interference at the graphene plane, which significantly enhances the localized electric field intensity. This enhanced field couples efficiently with the graphene plasmon modes, leading to improved plasmon resonance efficiency and enlarged in-plane temperature gradient in graphene, ultimately resulting in enhanced photovoltage response. Through comprehensive numerical simulations and experimental characterization, we systematically investigated the dependence of the PTE responsivity on reflector thickness and incident frequency. Using this detector with high responsivity, we further demonstrated the feasibility of terahertz imaging for distinguishing optically transparent liquids, highlighting its potential for high-performance, highly integrated terahertz sensing and imaging systems.

The excitation of plasmon resonance significantly enhances the electromagnetic field confinement and absorption within the graphene disk structures when the frequency of the incident terahertz wave matches the geometric parameters of the graphene array. To investigate the terahertz absorption characteristics of graphene microstructures on different substrate configurations, we employed FDTD Solutions to model a periodically repeating graphene disk array along the x- and y-directions on three representative substrate types for numerical simulations. When placed on a silicon substrate of infinite thickness with a 300 nm oxide layer, the absorption spectrum in the terahertz band exhibited a broad absorption peak (Fig. 1b). Upon modifying the substrate to a finite thickness high-resistivity silicon wafer, the absorption spectrum transformed into a Fabry–Pérot (FP) interference pattern modulated by the absorption profile of the graphene disk array. While the overall envelope of the curve remained consistent with the previous simulation, multiple absorption peaks emerged across the terahertz band (Fig. 1b). Subsequently, a perfect electric conductor (PEC) was introduced beneath the finite thickness substrate to achieve complete reflection of the electromagnetic wave. This resulted in a doubling of the maximum absorption value (Fig. 1c). We refer to this structure as interferometric enhancement of absorption (IEA), which is conceptually related to coherent absorption.[31]

The positions of the absorption peaks align with a dual beam interference model (Eq. 1-1), incorporating refractive indices of $n_{si} = 3.4$ and $n_{SiO_2} = 1.955$.

$$\frac{\lambda}{2} + \left(T_{Si} \times n_{si} + t_{SiO_2} \times n_{SiO_2}\right) \times 2 = m \times \lambda \qquad (1\text{-}1)$$

When the incident light is polarized along the x-direction, the electric field distribution in the xoz plane reveals interference fringes resulting from the interaction between the terahertz wave compressed within the substrate and the wave reflected by the PEC layer (Fig. 1d). It is evident that with an appropriate thickness of the substrate and dielectric layer, constructive interference can be achieved within the plane of the graphene structure,[32] thereby promoting in-plane plasmon polariton resonance (Fig. 1e). By varying the thickness of the silicon substrate, the absorption spectrum can be tailored for specific wavelengths (Fig. 1f). The uniform spacing between adjacent peaks indicates that under the influence of the IEA structure, the absorption at a single wavelength is predominantly determined by the substrate thickness.

Building on this foundation, we integrated the IEA structure with a graphene plasmon polariton atomic cavity (PPAC) photodetector. By coupling the energy from the Fabry-Pérot (FP) resonance in the substrate into the graphene atomic cavity, the absorption efficiency of the graphene microstructure for terahertz waves was significantly enhanced. Under terahertz illumination, the in-plane absorption heterogeneity of graphene, stemming from structural variations, induces a temperature gradient, which in turn generates a photothermal current via the Seebeck effect. Full channel modeling and simulation of the device show that under illumination at the same wavelength, the electric field distribution within the channel is strongly influenced by the IEA structure for a given substrate thickness (Fig. 1h). Moreover, the substrate thickness significantly affects the photovoltage response of the device. Due to the half-wave phase shift, the maximum response of the device with the IEA structure generally corresponds to the minimum response of the device without IEA (Fig. 1i).

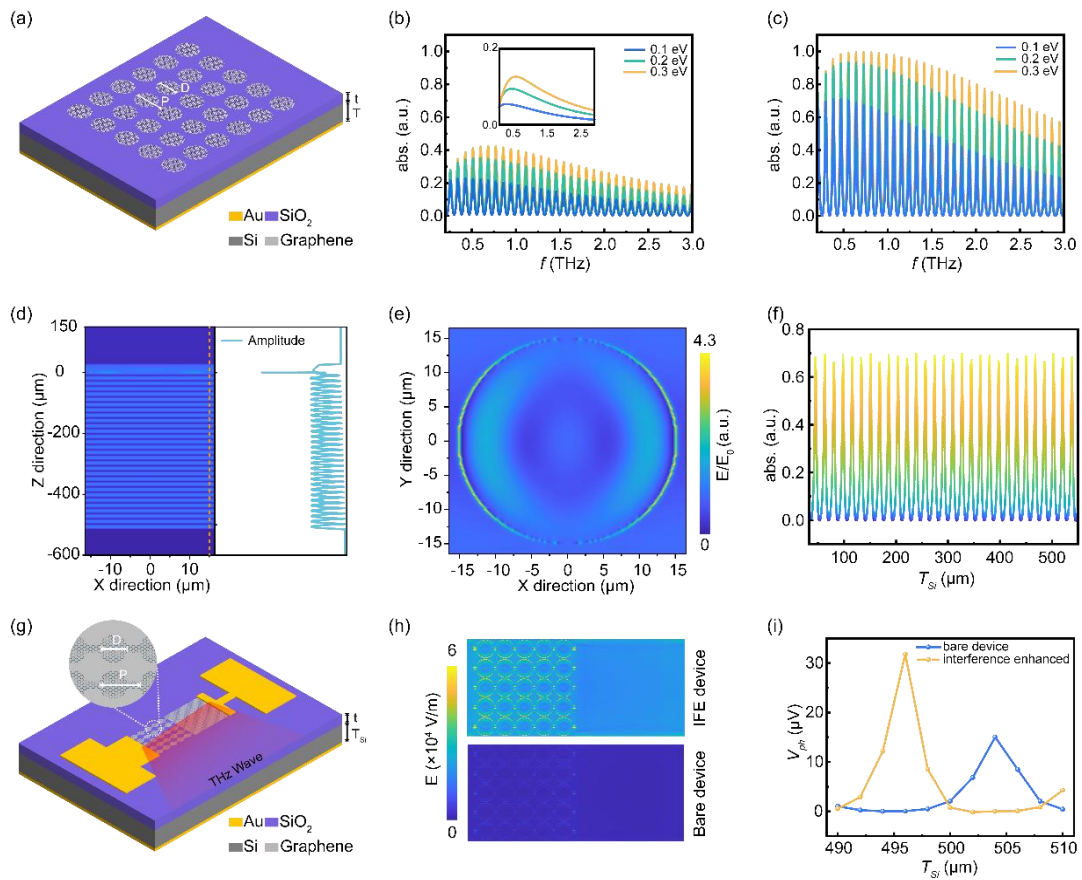

**Figure 1. (a)** Schematic of the simulation structure. The graphene disks have a diameter D=30 μm and a periodicity P=33 μm. **(b)** Absorption spectra of the structure with a finite-thickness Si substrate. The inset shows the absorption spectrum of the graphene disk array on an infinitely thick substrate. **(c)** Absorption spectrum of the disk array with a metal reflector layer. **(d)** Electric field distribution in the xoz plane. **(e)** In-plane (xoy) electric field distribution within the graphene layer. **(f)** The relationship between the absorption intensity of graphene disk arrays and substrate thickness. **(g)** Schematic diagram of the device structure. **(h)** Electric field distribution within the device channel for different substrate thicknesses. **(i)** Photovoltage across the channel as a function of substrate thickness.

Devices with and without the interferometric enhancement of absorption (IEA) structure were fabricated using microfabrication techniques (see Figure S1, Supporting Information). The I-V characteristics of the devices exhibited Ohmic contact behavior

(Fig. 2b), with a device resistance of approximately 2530 Ω. The introduction of the IEA structure did not alter the resistance distribution (Fig. 2f); both types of devices exhibited uniform and stable resistances within the same order of magnitude. As the incident terahertz power increased, the photoresponse of the device showed a linear growth trend (Fig. 2c), indicating good power dependent responsivity, which was measured to be 3.3 mV/W. When the polarization direction of the incident terahertz wave was varied, the photoresponse current remained largely polarization insensitive, consistent with the simulated absorption characteristics of the IEA graphene disk array (Fig. 2d). To more clearly compare the responses of devices with and without IEA, the switching characteristics of both devices are illustrated in Fig. 2e. It is evident that under the same illumination conditions, the IEA device exhibited a larger dynamic range. The average voltage responsivity increased from 0.068 mV/W to 2.119 mV/W, representing an improvement of nearly 30 times (Fig. 2f).

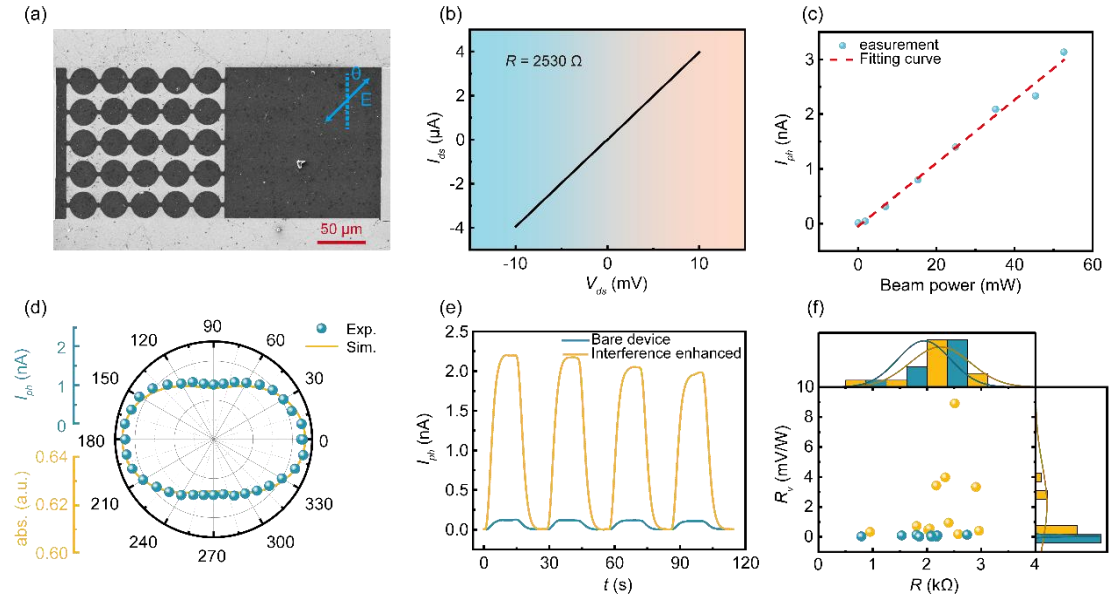

**Figure 2.** (a) SEM image of the device channel. (b) I–V characteristics. (c) Power-dependent photoresponse. (d) Polarization-dependent photoresponse. (e) Switching characteristics of devices with and without the IEA structure. (f) Statistical distribution of responsivity for devices with and without the IEA structure.

For the IEA device, the responsivity not only depends strongly on the substrate thickness but also varies with the frequency of the incident terahertz wave, due to wavelength dependent interference efficiency within the graphene layer. The device was tested using a frequency tunable electronic source, and the measured responsivity from 0.25 to 0.32 THz agreed well with simulation results (Fig. 3a), indicating a positive correlation between the IEA enhanced absorption and device responsivity.

The response bandwidth is another critical performance metric for detectors. Even when the chopper frequency was increased to its maximum operating frequency of 7.8 kHz, no significant decline in device response was observed (Fig. 3b). Given that the amplifier's effective bandwidth is 20 kHz, the response time of the device is estimated to be less than 130 μs. The noise power density of the device was characterized (Fig. 3c), revealing that thermal noise is the dominant noise source within the effective measurement range. Based on Eq. 1-5, [33] the noise equivalent power (NEP) was calculated to be $3.05\times10^{-10} W/Hz^{0.5}$. To evaluate response stability, the device was subjected to repeated light on/off cycles over 10 minutes. The resulting switching characteristics demonstrated excellent operational stability (Fig. 3d), with minor fluctuations at the peaks attributable to slight power variations in the laser output.

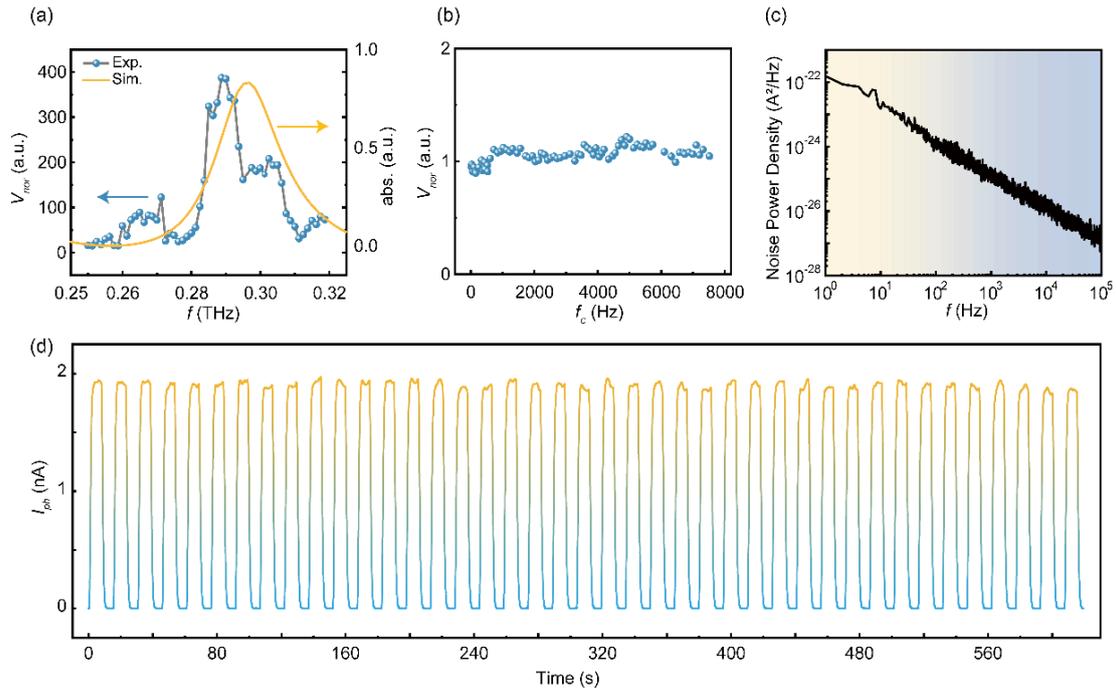

**Figure 3. (a) Frequency-dependent responsivity of the device. (b) Dependence of the photovoltage response on the chopper frequency. (c) Noise power spectral density of the device at zero bias. (d) Stability test over time.**

The IEA enhanced graphene photodetector, with its higher responsivity and dynamic range, offers robust performance for terahertz imaging applications, enabling non-destructive and non-contact imaging crucial for identifying materials with different compositions. As a proof of concept, the device was used to image two different liquids that are hidden. (Fig. 4a). A 55 μm thick filter paper was placed on a 1 mm thick, 3 cm × 2 cm quartz substrate (Fig. 4c). Approximately 100 μL of glycerol and paraffin oil were applied to the left and right sides, respectively, and spread evenly. The upper half of the sample was covered with a high resistivity silicon wafer (resistivity > 20,000

Ω·cm), with the boundary indicated by the orange dashed line in Fig. 4c.

The sample was mounted on a motorized translation stage, and the photocurrent from the detector was amplified, processed by a lock-in amplifier to suppress noise, and recorded by a computer. The normalized dual-focus scanning imaging result is shown in Fig. 4d. The sample was also examined using a transmission type terahertz time domain spectroscopy (THz-TDS) system[34] (Fig. 4b), with point-scan imaging used to record time domain spectra at each point. The time domain signals of glycerol with and without silicon cover are shown in Fig. 4e. The time shift between the two is mainly due to the additional optical path introduced by the silicon wafer, while the intensity difference arises from surface reflection of THz waves by silicon. Here, $t_1$ and $t_2$ correspond to the time offsets of the two main transmission peaks. The time domain scan image was Fourier transformed to extract frequency components from 0.1 to 3 THz (Fig. 4f). Figures 4g and 4h show scan images obtained by fixing the time delay at $t_1$ and $t_2$, respectively.

Analysis of the results indicates that both continuous wave THz dual-focus scanning imaging and broadband THz frequency domain point-scanning can distinguish between the two liquids and their covered regions, yet with significantly different temporal resolution. The acquisition time for Fig. 4d was approximately 5 hours, while that for Fig. 4f was about 46 hours. Furthermore, increasing the thickness of the silicon cover would further increase the time cost of acquiring full spectral information. In contrast, for continuous wave dual-focus scanning imaging, as long as the cover material (e.g., silicon) provides sufficient contrast in absorption or reflection between polar and non polar substances, the target can be effectively distinguished. It was observed that a single image could be obtained in about 1 hour when the time delay was fixed. Although combining multiple images with fixed time delays can also identify the liquid distribution in both covered and uncovered areas, samples with significantly different refractive indices in a single measurement can degrade the imaging quality due to the time shift of the signals.

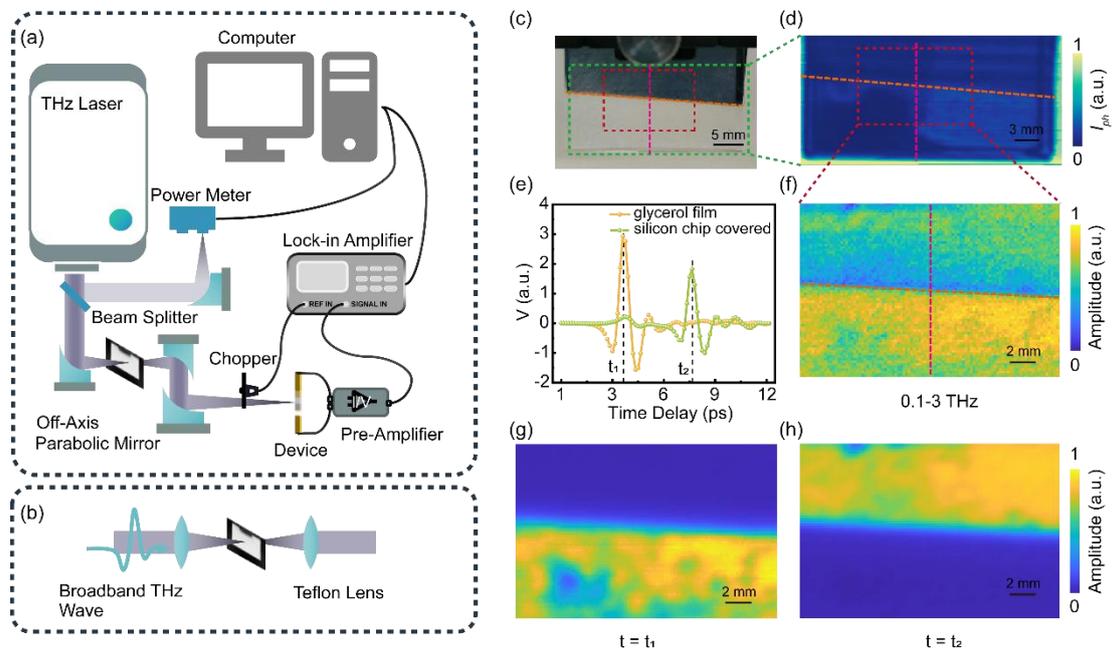

**Figure 4. (a) Schematic of the photocurrent dual-focus scanning imaging setup. (b) Schematic of the terahertz time-domain spectroscopy (THz-TDS) point scan imaging system. (c) Photograph of the imaged sample: a liquid film partially covered by a silicon wafer. The magenta dashed line indicates the boundary between the glycerol film (left) and the paraffin oil film (right). (d) Image of the sample obtained under continuous-wave terahertz illumination. (e) Time-domain signals of the glycerol film with and without silicon cover. (f) Frequency-domain image of the sample under broadband terahertz illumination. (g, h) Images of the sample at time delays $t = t_1$ and $t = t_2$,**

Through imaging experiments, we demonstrate that the device incorporating the IEA structure enables effective discrimination and imaging of concealed polar and non polar liquids. Compared to terahertz time domain spectroscopy (THz-TDS) which can also distinguish between the two liquids, however, our approach significantly improves temporal resolution. Owing to its high dynamic response range, the IEA based device clearly identifies the distribution of liquid films both with and without silicon coverage, effectively collapsing both the refractive index and absorption information of the sample into a one dimensional photocurrent response.

CONCLUSION

This study successfully designed and validated a terahertz photodetector based on the synergistic enhancement between a back side metal reflector (IEA structure) and graphene plasmon polariton atomic cavity (PPAC) based photodetector. By tailoring the spacing between the metal reflector and the graphene micro structure, coherent superposition of incident and reflected terahertz waves was achieved, significantly enhancing the localized electric field and optical absorption efficiency in graphene, thereby greatly improving the photothermoelectric response. Compared with the bare device, this structure increased the responsivity of the device by approximately 30 times at 2.52 THz, reaching 2.119 mV/W, with a response time below 130 μs, while maintaining polarization insensitive characteristics and stable room temperature operation. Additionally, the detector exhibited a high dynamic range and excellent signal to- noise ratio in imaging applications, effectively distinguishing between concealed polar and non-polar liquids with significantly reduced imaging time compared to traditional terahertz time domain spectroscopy systems. This study not only provides an effective and process compatible structural strategy for high performance, room temperature, and compact terahertz detectors but also demonstrates its potential for rapid terahertz imaging and non-destructive testing applications. Besides, the devices presented in this work were fabricated using CVD grown graphene, which offers promising potential for future high-density integration of such detector arrays.

METHODS

Sample Preparation

A large area monolayer graphene film (purchased from XFNANO) was first transferred onto the oxidized side of a double side polished substrate via a wet transfer method. The graphene was then patterned using maskless lithography (Tuotuo Technology) followed by oxygen plasma etching. Subsequently, another round of maskless lithography was performed, and electrodes were fabricated by thermal evaporation of 10 nm Cr and 70 nm Au. The electrode coated surface was spin coated with photoresist for protection, while selected regions on the back side of the substrate were covered and deposited with 10 nm Cr and 100 nm Au via evaporation. Finally, the device was completed through a lift-off process.

Measurement

A continuous wave terahertz laser (FIRL 100) operating at 2.52 THz was used for device response and photocurrent dual-focus scanning imaging. The beam, with a spot diameter of 1 mm, was modulated into a quasi square wave using an optical chopper before illuminating the device. The signal from the device electrodes was transmitted via coaxial cable to a low noise transimpedance amplifier (DLPCA-200). The amplified voltage signal was then processed by a lock-in amplifier (OE1201) to extract the photoresponse by noise suppression.

Responsivity Calculation

The voltage signal $V_{lock}$ measured by the lock-in amplifier is related to the photocurrent $I_{ph}$ by:

$$I_{ph} = \frac{2\pi\sqrt{2}V_{lock}}{4G} \tag{1-2}$$

where $G$ is the gain of the preamplifier. Since no bias voltage was applied, the voltage responsivity $R_v = \frac{I_{ph} \times R}{P_{eff}}$ (in mV/W) is defined as:

$$R_v = \frac{I_{ph} \times R}{P_{eff}} \tag{1-3}$$

$$P_{eff} = \frac{S_D}{S_B} \times P_B \tag{1-4}$$

Here, $I_{ph}$ is the photocurrent, $R$ is the resistance of the detector, $P_{eff}$ is the effective optical power, $S_D$ is the area of the device channel, $S_B$ is the beam spot area, and $P_B$ is the power of the terahertz beam.

NEP Calculation

$$NEP = \frac{i_{noise}}{R_v} \sqrt{B} \tag{1-5}$$

where $i_{noise}$ is the total noise current and $B$ is the bandwidth.

ASSOCIATED CONTENT

**Supporting Information**.

The following files are available free of charge.

Figures S1–S7, details of theoretical modeling, experiment and data analysis. (PDF)

AUTHOR INFORMATION

**Corresponding Author**

**Huanjun Chen** — State Key Laboratory of Optoelectronic Materials and Technologies, Guangdong Province Key Laboratory of Display Material and Technology, School of Electronics and Information Technology, Sun Yat-sen University, Guangzhou 510275, China; orcid.org/0000-0003-

4699- 009X; Email: chenhj8@mail.sysu.edu.cn

**Author**

**Runli Li** — State Key Laboratory of Optoelectronic Materials and Technologies, Guangdong Province Key Laboratory of Display Material and Technology, School of Electronics and Information Technology, Sun Yat-sen University, Guangzhou 510275, China

**Shaojing Liu** — State Key Laboratory of Optoelectronic Materials and Technologies, Guangdong Province Key Laboratory of Display Material and Technology, School of Electronics and Information Technology, Sun Yat-sen University, Guangzhou 510275, China

**Ximiao Wang** — State Key Laboratory of Optoelectronic Materials and Technologies, Guangdong Province Key Laboratory of Display Material and Technology, School of Electronics and Information Technology, Sun Yat-sen University, Guangzhou 510275, China

**Hongjia Zhu** — State Key Laboratory of Optoelectronic Materials and Technologies, Guangdong Province Key Laboratory of Display Material and Technology, School of Electronics and Information Technology, Sun Yat-sen University, Guangzhou 510275, China

**Yongsheng Zhu** — State Key Laboratory of Optoelectronic Materials and Technologies, Guangdong Province Key Laboratory of Display Material and Technology, School of Electronics and Information Technology, Sun Yat-sen University, Guangzhou 510275, China

**Shangdong Li** — State Key Laboratory of Optoelectronic Materials and Technologies, Guangdong Province Key Laboratory of Display Material and Technology, School of

Electronics and Information Technology, Sun Yat-sen University, Guangzhou 510275, China

**Author Contributions**

H.C. conceived and designed the experiments. Sample fabrication was performed by R. L.. M.W., Y.Z. and S.L programmed all equipment. R.L., S.L. and H.C. analyzed the results. H.Z. and R.L. performed the experiments and numerical simulations. R.L. and S.L. wrote the manuscript. †R.L. and S.L. contributed equally. All authors have given approval to the final version of the manuscript.

**Notes**

The authors declare no competing financial interest.

ACKNOWLEDGMENT

The authors acknowledge support from the National Key Basic Research Program of China (grant nos. 2024YFA1208500, 2024YFA1208501, and 2025YFA1213200), the National Natural Science Foundation of China (grant No. 92463308), Guangdong Basic and Applied Basic Research Foundation (grant No. 2023A1515011876), and China Postdoctoral Science Foundation (grant No. 2025M780806).

ABBREVIATIONS

THz, terahertz; PTE, photothermoelectric; IEA, interferometric enhancement of absorption; PPAC, plasmon polariton atomic cavity; GPP, graphene plasmon polariton; LSPR, localized surface plasmon resonance; FP, Fabry–Pérot; PEC, perfect electric conductor; FDTD, finite-difference time-domain; CW, continuous wave; NEP, noise equivalent power; SEM, scanning electron microscopy; THz-TDS, terahertz time-domain spectroscopy; CVD, chemical vapor deposition.